\documentclass[aps,prd,twocolumn,superscriptaddress,showpacs]{revtex4}

\usepackage{graphicx}
\usepackage{color}
\usepackage{amssymb}
\usepackage{amsmath}

\begin{document}

\title{Beam-Spin Asymmetries in the Azimuthal Distribution of Pion Electroproduction}


\def\groupargonne{\affiliation{Physics Division, Argonne National Laboratory, Argonne, Illinois 60439-4843, USA}}
\def\groupbari{\affiliation{Istituto Nazionale di Fisica Nucleare, Sezione di Bari, 70124 Bari, Italy}}
\def\groupbeijing{\affiliation{School of Physics, Peking University, Beijing 100871, China}}
\def\groupchina{\affiliation{Department of Modern Physics, University of Science and Technology of China, Hefei, Anhui 230026, China}}
\def\groupcolorado{\affiliation{Nuclear Physics Laboratory, University of Colorado, Boulder, Colorado 80309-0390, USA}}
\def\groupdesy{\affiliation{DESY, 22603 Hamburg, Germany}}
\def\groupzeuthen{\affiliation{DESY, 15738 Zeuthen, Germany}}
\def\groupdubna{\affiliation{Joint Institute for Nuclear Research, 141980 Dubna, Russia}}
\def\grouperlangen{\affiliation{Physikalisches Institut, Universit\"at Erlangen-N\"urnberg, 91058 Erlangen, Germany}}
\def\groupferrara{\affiliation{Istituto Nazionale di Fisica Nucleare, Sezione di Ferrara and Dipartimento di Fisica, Universit\`a di Ferrara, 44100 Ferrara, Italy}}
\def\groupfrascati{\affiliation{Istituto Nazionale di Fisica Nucleare, Laboratori Nazionali di Frascati, 00044 Frascati, Italy}}
\def\groupgent{\affiliation{Department of Subatomic and Radiation Physics, University of Gent, 9000 Gent, Belgium}}
\def\groupgiessen{\affiliation{Physikalisches Institut, Universit\"at Gie{\ss}en, 35392 Gie{\ss}en, Germany}}
\def\groupglasgow{\affiliation{Department of Physics and Astronomy, University of Glasgow, Glasgow G12 8QQ, United Kingdom}}
\def\groupillinois{\affiliation{Department of Physics, University of Illinois, Urbana, Illinois 61801-3080, USA}}
\def\groupmichigan{\affiliation{Randall Laboratory of Physics, University of Michigan, Ann Arbor, Michigan 48109-1040, USA }}
\def\groupmoscow{\affiliation{Lebedev Physical Institute, 117924 Moscow, Russia}}
\def\groupnikhef{\affiliation{Nationaal Instituut voor Kernfysica en Hoge-Energiefysica (NIKHEF), 1009 DB Amsterdam, The Netherlands}}
\def\groupstpetersburg{\affiliation{Petersburg Nuclear Physics Institute, St. Petersburg, Gatchina, 188350 Russia}}
\def\groupprotvino{\affiliation{Institute for High Energy Physics, Protvino, Moscow region, 142281 Russia}}
\def\groupregensburg{\affiliation{Institut f\"ur Theoretische Physik, Universit\"at Regensburg, 93040 Regensburg, Germany}}
\def\grouprome{\affiliation{Istituto Nazionale di Fisica Nucleare, Sezione Roma 1, Gruppo Sanit\`a and Physics Laboratory, Istituto Superiore di Sanit\`a, 00161 Roma, Italy}}
\def\grouptriumf{\affiliation{TRIUMF, Vancouver, British Columbia V6T 2A3, Canada}}
\def\grouptokyo{\affiliation{Department of Physics, Tokyo Institute of Technology, Tokyo 152, Japan}}
\def\groupamsterdam{\affiliation{Department of Physics and Astronomy, Vrije Universiteit, 1081 HV Amsterdam, The Netherlands}}
\def\groupwarsaw{\affiliation{Andrzej Soltan Institute for Nuclear Studies, 00-689 Warsaw, Poland}}
\def\groupyerevan{\affiliation{Yerevan Physics Institute, 375036 Yerevan, Armenia}}
\def\groupnone{\noaffiliation}


\groupargonne
\groupbari
\groupbeijing
\groupchina
\groupcolorado
\groupdesy
\groupzeuthen
\groupdubna
\grouperlangen
\groupferrara
\groupfrascati
\groupgent
\groupgiessen
\groupglasgow
\groupillinois
\groupmichigan
\groupmoscow
\groupnikhef
\groupstpetersburg
\groupprotvino
\groupregensburg
\grouprome
\grouptriumf
\grouptokyo
\groupamsterdam
\groupwarsaw
\groupyerevan


\author{A.~Airapetian}  \groupmichigan
\author{Z.~Akopov}  \groupyerevan
\author{M.~Amarian}  \groupzeuthen \groupyerevan
\author{A.~Andrus}  \groupillinois
\author{E.C.~Aschenauer}  \groupzeuthen
\author{W.~Augustyniak}  \groupwarsaw
\author{H.~Avakian\footnote[1]{Present address: Thomas Jefferson National Accelerator Facility, Newport News, Virginia 23606, USA}}  \groupfrascati
\author{R.~Avakian}  \groupyerevan
\author{A.~Avetissian}  \groupyerevan
\author{E.~Avetisyan}  \groupfrascati \groupyerevan
% ...

\collaboration{The HERMES Collaboration} \noaffiliation

\date{\today}
\begin{abstract}
{A measurement of the beam-spin asymmetry
 in the azimuthal distribution of pions produced
in semi-inclusive deep-inelastic
scattering off protons is presented. The measurement was performed
using the \mbox{HERMES} spectrometer with a hydrogen gas
target and the longitudinally polarized 27.6~GeV positron beam of HERA.
The  sinusoidal amplitude of the dependence of the asymmetry  
on the angle $\phi$ of the hadron production plane around the 
virtual photon direction relative to the lepton scattering plane
was measured for 
$\pi^+,\pi^-$ and $\pi^0$ mesons. The dependence of
 this amplitude on the Bjorken scaling variable and on the 
pion fractional energy and transverse 
momentum is presented. The results are compared to theoretical model calculations.}
\end{abstract}

\pacs{13.60.-r; 13.87.Fh; 13.88.+e; 14.20.Dh; 24.85.+p}

\maketitle


Single-spin asymmetries (SSA) in semi-inclusive deep-inelastic 
scattering (SIDIS) are known as a powerful tool to probe the 
partonic structure of the nucleon. 
If the orbital motion of the quarks is neglected, the
structure of nucleon can be described in the leading twist by 3 parton distribution functions (PDF)
defining the momentum $f_1$, helicity $g_1$, and transversity $h_1$
distributions.
The observation of different azimuthal asymmetries and, in particular 
SSAs were an indication of a more complex inner structure of the nucleon. 
Those effects were recognized to be 
due to
correlations of spin  and transverse momentum of quarks and/or hadrons and
appear as 
moments of the azimuthal angle between scattering and production planes. 
The $\sin\phi$ azimuthal moments contain
contributions from different chiral-odd
and/or na\"ive {\it time-reversal-odd} (T-odd) distribution and 
fragmentation functions, arising from interference of wave functions for 
different orbital angular momentum states and final state interactions \cite{KOTZ,ANS,TANG,BELJI,BEL,JIMA,JAFFE,BOER,TODD,METZ}. 
Experimentally correlations between spin and transverse momentum either in the initial 
target nucleon (e.g. the Sivers mechanism \cite{Sivers}) or in the 
fragmentation process (e.g. the Collins mechanism \cite{COL}) result in an asymmetry of the distribution of hadrons around the virtual photon direction. 
Asymmetries are attractive observables as they are expected to be
less sensitive to a number of higher order corrections
than cross sections measured in SIDIS \cite{BACC}.
SSAs in production of pseudoscalar mesons were studied with both
longitudinally  \cite{AULL,AULPI0,AUL} and transversely \cite{TRANS,COMPASS} polarized
targets. Recently, SSAs have been observed also in SIDIS with longitudinal polarized beams 
and unpolarized targets (in the following refered to as beam SSAs) \cite{AULL,JLAB}.

Under the assumption of factorization \cite{FACT,Ji:2004xq},
the general expression for the  SIDIS cross section $\sigma$ can be given as 
convolutions of distribution functions $f^{H\rightarrow q}$ (DF), elementary 
hard scattering  cross section $\sigma^{eq\rightarrow eq}$, and fragmentation functions $D^{q\rightarrow h}$ (FF),
\begin{equation}
\sigma^{eH\rightarrow ehX} = {f^{H\rightarrow q}} \otimes 
\sigma^{eq\rightarrow eq} \otimes D^{q\rightarrow h}
\label{conv} .
\end{equation}

\begin{figure}[tb]
%
%
%
\includegraphics[width=6.5cm]{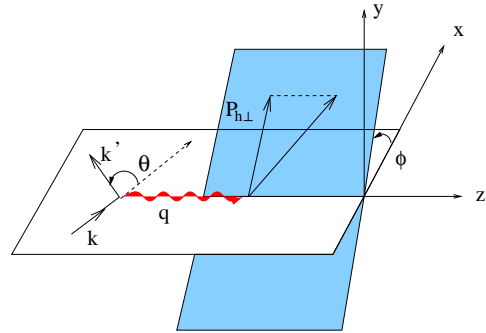}
\caption[*]{Definition of kinematic planes for semi-inclusive deep-inelastic scattering.
}
\label{kinem}
\end{figure}
\vspace{0.5cm}

\noindent In the particular case of a longitudinally polarized beam (L) and an unpolarized target 
(U), and in the 
limit of massless quarks, the differential polarized cross section $\sigma_{LU}$
 can be written as a sum of convolutions
of twist-2 and twist-3 functions \cite{BACNEW,BAC}
\begin{equation} \begin{split}
\label{siglu}
\frac{d^5\sigma_{LU}}{dxdydzd\phi d P_{h\perp}^2} & \propto
\lambda_e  \sin\phi\,y \sqrt{1-y} \frac{M}{Q} \times \\
  & \int d^2 \! \boldsymbol{p_T} d^2 \boldsymbol{k_T} \,
         \delta^{(2)}\Bigl(\boldsymbol{p_T}
         -\frac{\boldsymbol{P_{h \perp}}}{z}-\boldsymbol{k_T}\Bigr) \\
  & \biggl\{ \frac{\boldsymbol{\hat{P}_{h \perp}}\cdot
         \boldsymbol{p_T}}{M}
         \left[\frac{M_h}{Mz}\, h_1^{\perp} E +x\, g^\perp D_1\right] - \\
  &  \frac{\boldsymbol{\hat{P}_{h \perp}}
         \cdot \boldsymbol{k_T}}{M_h}
         \left[\frac{M_h}{Mz}\,f_1\,G^\perp + x\, e H_1^{\perp} \right] \biggr\}.
\end{split}\end{equation}
In Eq.~\ref{siglu},  $\lambda_e$ is the lepton helicity,
$M$ and $M_h$ are the nucleon and hadron masses, 
$-Q^2$ is the 4-momentum transfer squared, $\boldsymbol{P_{h \perp}}$ is the transverse momentum of the 
detected hadron with 
$\boldsymbol{\hat{P}_{h \perp}}=\boldsymbol{P_{h \perp}}/|\boldsymbol{P_{h \perp}}|$,
$\boldsymbol{p_T}$ ($\boldsymbol{k_T}$) is the intrinsic 
quark transverse momentum in the generic distribution function $f$ (fragmentation function
$D$), $E$($E^\prime$) and $E_h$ are the energies of the incoming(scattered) lepton and the hadron produced, $\nu$ is the energy of the virtual photon 
and the azimuthal angle $\phi$ is defined as the angle
between the lepton-scattering and hadron-production planes 
according to the Trento convention \cite{TRENTO}. See Fig.~\ref{kinem} for the 
definition of the kinematic planes, where $k$~($k^\prime$) is the four-momentum of the incoming (scattered) lepton and $q$ that of the virtual photon. 
In Eq.~\ref{siglu}, a charge-weighted sum over quark and antiquark flavours is implicit. The quantities $f_1$ and $D_1$ are the
 twist-2 DF and FF,
which appear in the unpolarized cross-section when integrated over $\phi$
\begin{equation}
\frac{d^3\sigma_{UU}}{dxdydz}\propto (1-y+y^2/2)
f_1(x) D_1(z) . 
\label{sigmauu}
\end{equation}
In Eq.~\ref{siglu}, $e$ is a chiral-odd twist-3 unpolarized DF \cite{LM} that can be related to
  the pion-nucleon $\sigma$-term, which in its turn is related 
  to the strangeness content of the nucleon \cite{SCHWEITZER}. 
  The T-odd DF $e$ is convolved with the twist-2 Collins FF $H_1^{\perp}$,
  which also appears in longitudinal and 
transverse target-spin asymmetries \cite{TRANS}.
Another contribution is given by the twist-2 DF
$h_1^\perp$ (Boer-Mulders DF \cite{BOER}), which is interpreted as 
representing the correlation between the transverse spin and intrinsic 
transverse momentum of a quark in an unpolarized nucleon.
Here it is convolved with the twist-3
chiral-odd FF $E$ \cite{KOGAN,YUAN}. The remaining terms 
contain the twist-3 DF $g^\perp$ and FF $G^\perp$  convolved with
the unpolarized FF and DF, respectively.
The DF $g^\perp$ can be directly accessed
through a measurement of the beam-spin asymmetry in jet 
production \cite{AFANAS,BAC}.
Since the beam SSA has no leading-twist contribution (cf. Eq.~\ref{siglu}) it is 
expected to be accessible only at moderate values of $Q^2$.

In this paper we present a measurement of the beam SSA
for charged and neutral pions produced in SIDIS at \mbox{HERMES} during the years 1996-2000.
The  results presented supercede a previous HERMES measurement of the beam SSA for $\pi^+$ \cite{AULL} by almost doubling the statistics, which allowed the extraction of the kinematic dependences of the beam SSA on $z$, $x$ and $P_{h\perp}$ and the addition of a measurement for $\pi^-$ and $\pi^0$ mesons. 
The experiment used the polarized positron beam of the HERA accelerator and a
hydrogen gas target. 
Positrons with an energy of $27.6$~GeV were scattered off hydrogen nuclei in an atomic gas target \cite{TARG}. The beam was polarized in the transverse direction due to the Sokolov-Ternov effect \cite{SOKTER}. Longitudinal orientation of the beam spin was obtained by using a pair of spin rotators located before and behind the interaction region of HERMES. The beam helicity was flipped every few months. The beam polarization was measured by two independent HERA polarimeters \cite{BPOL} and had an average value of $0.53$ with a fractional systematic uncertainty of $2.9\%$.
The target mode was either unpolarized or longitudinally polarized with a fast (90~s) helicity flip. The resulting target polarization in the analyzed sample was
$-1.3\cdot10^{-4}$, which is consistent with zero and was safely ignored in the analysis.
The scattered positrons and associated hadrons were 
detected by the \mbox{HERMES} spectrometer \cite{SPEC}.
 Positrons were distinguished from hadrons
by the use of a set of particle identification detectors: a transition-radiation detector,
a preshower radiator/hodoscope, 
a threshold \v Cerenkov detector (upgraded to a RICH detector \cite{RICH} in 1998) 
and 
an electromagnetic calorimeter \cite{CALO}. The average positron identification efficiency exceeded 98\% with
a hadron contamination in the positron sample below 1\%.
Several kinematic requirements were imposed on the scattered positron, namely
\mbox{$1<Q^2<15$ GeV$^2$}, 
$0.023<x<0.4$, \mbox{$W^2>4$ GeV$^2$}, $y<0.85$.

For identification of charged pions the \v Cerenkov and RICH detectors were used 
during the corresponding data taking periods. To assure reliable identification
of pions in both detectors the momentum range of $4.5<P<13.5$ GeV was chosen.

Neutral pions were identified requiring two neutral
clusters in the electromagnetic calorimeter with energies above a threshold of 1 GeV. 
A peak around the $\pi^0$
mass of $0.135$ GeV with a resolution of about $0.012$ GeV was clearly
observed in the 
invariant mass $M_{\gamma\gamma}$ distribution of all photon pairs. Neutral
pions 
were selected requiring \mbox{$0.11<M_{\gamma\gamma}<0.16$ GeV}.
By fitting the $M_{\gamma\gamma}$ distribution with a Gaussian plus 
a polynomial of third order,
the combinatorial background was estimated to be up to $35\%$ in the 
lower-$z$ region and negligible ($<5\%$) for $z>0.5$. 
A momentum cut of $P>2$ GeV was used to reduce the low-energetic combinatorial background. The spatial granularity of the electromagnetic calorimeter imposes an upper limit of about $P<15$ GeV on the $\pi^0$ momenta.

\begin{figure}[t]
\vspace*{6cm}
\begin{center}
\includegraphics{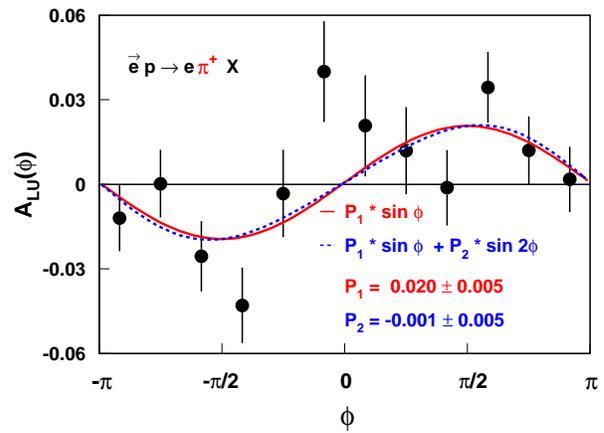}
\caption{Beam SSA as a function of $\phi$ for $\pi^+$ electroproduction at 
mid-$z$ range. 
The solid curve
represents a $\sin\phi$ fit, and the dashed one includes also the $\sin2\phi$
harmonic. Only statistical errors are shown.}
\label{aphi}
\end{center}
\end{figure}

The dependence of the cross-section asymmetry on the azimuthal angle $\phi$  
was extracted as
\begin{equation}
  \label{asy}
 A_{LU}(\phi)=\frac{1}{|P_B|}\frac{\stackrel{\rightarrow}{N}(\phi) -
\stackrel{\leftarrow}{N}(\phi)}{\stackrel{\rightarrow}{N}(\phi)
+\stackrel{\leftarrow}{N}(\phi)},
\end{equation}
where $|P_B|$ is the average absolute luminosity-weighted beam polarization, $\rightarrow(\leftarrow)$ denotes  positive (negative) 
helicity of the beam
and $\stackrel{\rightarrow}{N} (\stackrel{\leftarrow}{N})$ 
is the number of selected events with a 
detected pion for each beam spin state normalized to DIS. 
The cross-section asymmetry dependence on $\phi$ for $\pi^+$ mesons is shown
in Fig.~2 at mid-$z$ range with 
mean kinematic values
\mbox{($0.5<z<0.8$)}:
\mbox{$\langle z\rangle=0.62$},
\mbox{$\langle x\rangle=0.10$}, \mbox{$\langle Q^2\rangle=2.55$~GeV$^2$}, 
and \mbox{$\langle P_{h\perp} \rangle=0.45$~GeV}.
Fourier amplitudes were extracted with
fits using the functions 
$p_1\cdot\sin\phi$ and $p_1\cdot\sin\phi + p_2\cdot\sin2\phi$, also shown in
the figure.
For both fits the resulting asymmetry amplitude $A_{LU}^{\sin\phi}$
equals  $0.020\pm0.005$. The amplitude $A_{LU}^{\sin2\phi}$ is compatible
with zero.

\begin{figure}[t]
\vspace*{11.0cm}
\begin{center}
\includegraphics{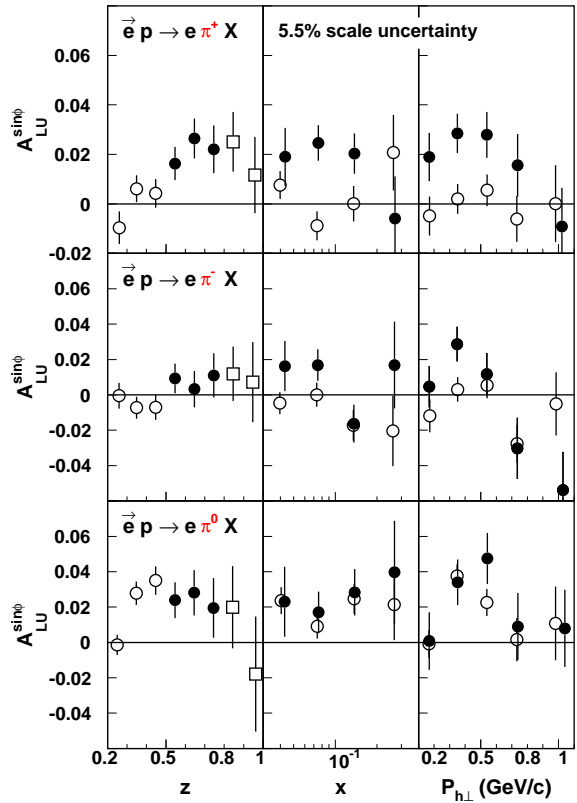}
\caption[*]{ Dependence of the beam SSA on $z$, $x$, and $P_{h\perp}$.
Results for the $x$ and $P_{h\perp}$ dependences are presented
separately for the  low-$z$ ($0.2<z<0.5$) and mid-$z$ ($0.5<z<0.8$) regions,
indicated by  open and full circles, respectively. For the high-$z$ ($0.8<z<1$, open squares)
region only the $z$ dependence is pictured. 
The error bars represent the statistical uncertainty.
An additional $5.5\%$ fractional scale uncertainty 
is due to the systematic uncertainty in the 
beam polarization measurement. Total 
systematic uncertainties do not exceed $0.005$.
 }
\label{alupi}
\end{center}
\end{figure}

The amplitudes $A_{LU}^{\sin\phi}$ for all pions are shown in Fig.~\ref{alupi}
and Table~\ref{aluzdata} as a function of $z$, $x$, and 
$P_{h\perp}$. Three regions in $z$ are considered: the low-$z$ 
($0.2<z<0.5$), the mid-$z$ ($0.5<z<0.8$),
and the high-$z$ region ($0.8<z<1$). In the latter region the contributions of 
exclusive processes become sizeable.
The distributions of $A_{LU}^{\sin\phi}$ 
in $x$ and $P_{h\perp}$ are extracted for the low- and mid-$z$ 
regions separately and presented as open and full circles, respectively.

The sources of systematic uncertainties are the
beam polarization measurement with an average relative uncertainty on 
the asymmetry 
of $5.5\%$, 
radiative processes, acceptance effects, asymmetry amplitude extraction method
and the hadron 
identification efficiency. The combined systematic uncertainty, excluding the 
contribution from the beam polarization measurement,
was evaluated by Monte Carlo studies and found to be less than 0.005 in total. 

For the determination of the $\pi^0$ asymmetries, the asymmetry of the combinatorial 
background $A_{LU}^{bg}$ was measured outside the mass window 
of the $\pi^0$ peak and found to equal about $0.03$ on average. Since the contribution from the combinatorial background is negligible in the mid- and high-$z$ regions, a systematic uncertainty due to combinatorial background subtraction adds to the total systematic uncertainty in the low-$z$ region reaching $0.007$.
The measured asymmetry $A_{LU}^{meas}$ was  corrected in each kinematic bin
using the equation

\begin{eqnarray}
A_{LU}^{corr}=\frac{A_{LU}^{meas}N_{meas}-A_{LU}^{bg}N_{bg}}{N_{meas}-N_{bg}} 
\label{subtract} ,
\end{eqnarray}
where  $N_{meas}$ and $N_{bg}$ are the number of photon pairs in
the region considered in the invariant mass distribution
and uncorrelated photon pairs, respectively.

The amplitude $A_{LU}^{\sin\phi}$ for $\pi^+$ mesons is found to be positive
on average. It is compatible with zero in the low-$z$ region 
and exhibits a rise to values of about
$0.02$ for increasing $z$. The $x$ and $P_{h\perp}$ dependences of
the amplitude are consistent with zero in the low-$z$ region 
whereas in the mid-$z$ region they decrease at large $x$ and
$P_{h\perp}$. 
        
The amplitude $A_{LU}^{\sin\phi}$ for $\pi^-$ mesons is consistent 
with zero in the whole $z$ range, with fluctuations around zero
in the $x$ and $P_{h\perp}$ distributions. 

The asymmetry for $\pi^0$ mesons is positive and of the order of about $0.03$
in the whole $z$ range except in the highest and lowest bins where the
asymmetry  
is compatible with zero. 
The dependence of the 
asymmetry amplitude on $x$ is weak 
while for $P_{h\perp}$ the amplitude decreases
at higher $P_{h\perp}$.

\begin{figure}[t]
\vspace*{7.5cm}
\begin{center}
\includegraphics{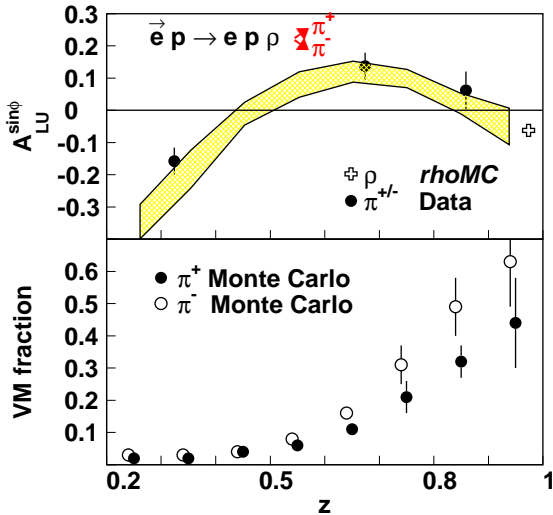}
\caption[*]{Top panel: amplitude $A_{LU}^{\sin\phi}$ for $\pi^+$ mesons
  originating from $\rho^0$ meson decays, 
obtained with Monte Carlo (band) and data (full circles).
The open cross displays the asymmetry for the $\rho^0$ itself (Monte Carlo).
Bottom panel: the fraction of pions in the SIDIS sample originating from
VM decays.
}
\label{RHOCONT}
\end{center}
\end{figure}

Semi-inclusive pion production 
($ep\rightarrow e^\prime \pi X$)
 with an underlying mechanism of quark fragmentation
is diluted by exclusive vector meson (VM)
production which can contribute significantly in certain kinematic regions at HERMES.
In Fig.~\ref{RHOCONT} (lower panel) the relative contribution of exclusive 
VM production in the 
semi-inclusive pion sample is shown as obtained
with the {\sc Pythia} Monte-Carlo generator tuned for \mbox{HERMES} kinematics
\cite{PATTY}.  
The VM contribution increases with $z$ from about $4\%$ in the lowest $z$-bin
to approximately $40\% (60\%) $ in the highest $z$-bin for the $\pi^+(\pi^-)$ mesons.
The $\pi^0$ meson sample is less contaminated with VM decay products since the 
$\omega$ meson production rate is very small at \mbox{HERMES} energies;
hence its 
contribution to the $\pi^0$ sample does not exceed $5\%$ in the  
kinematic range considered.  

To assess the effect of the exclusive processes, the amplitude
$A_{LU}^{\sin\phi}$ has been extracted for pions identified as decay products
of exclusive $\rho^0$ mesons in the data, and compared with a Monte-Carlo simulation
based on the VMD model and spin-density matrix elements extracted from HERMES data \cite{RHOMC}.
The asymmetry as a function of $z$ extracted for $\pi^+$ from the Monte Carlo
and the data  
is presented in Fig.~\ref{RHOCONT} (upper panel).
Due to the symmetric decay of the $\rho^0$ mesons the 
asymmetry amplitudes of $\pi^-$ and $\pi^+$ mesons are identical.
The agreement between data and Monte Carlo within the available statistical accuracy 
is clearly visible.

\begin{figure}[t]
\vspace*{11.0cm}
\begin{center}
\includegraphics{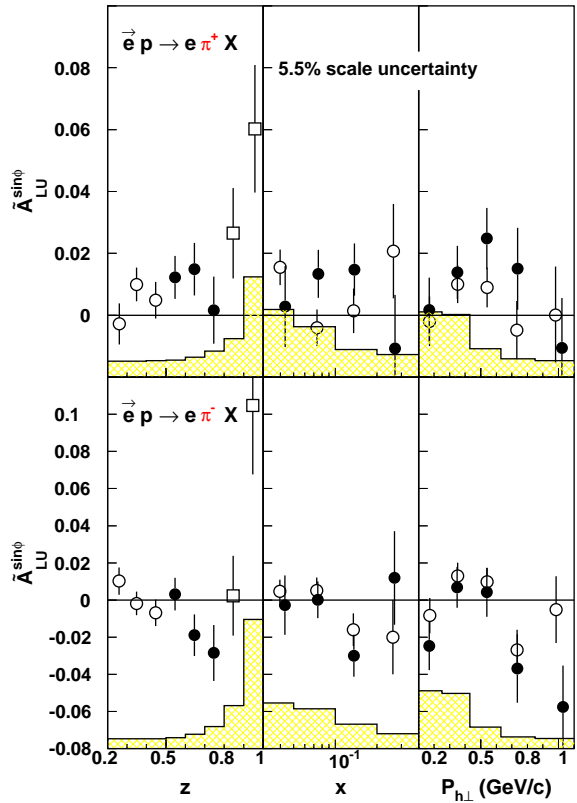}
\caption[*]{Dependence of $\widetilde{A}_{LU}^{\sin\phi}$ 
on $z$, $x$ and $P_{h\perp}$
for charged pions. The contribution from VM decays has been determined 
from a Monte-Carlo
simulation and subsequently subtracted from the asymmetries.
The measurement of the $x$ and $P_{h\perp}$ dependences is
made separately for low ($0.2<z<0.5$) and middle ($0.5<z<0.8$) $z$-ranges
(indicated by open and full circles, respectively). The error band indicates 
the uncertainties from {\sc Pythia} and {\sc rhoMC}. }
\label{RHOSUB}
\end{center}
\end{figure}

The information obtained from the Monte-Carlo simulation was used 
to subtract the contribution from exclusive VM production 
to the measured asymmetry amplitudes $A_{LU}^{\sin\phi}$.
The subtraction was performed similarly to the combinatorial 
background correction in the $\pi^0$ meson sample (cf.~Eq.~\ref{subtract}). 
The corrected 
asymmetry $\widetilde{A}_{LU}^{\sin\phi}$ is presented in Fig.~\ref{RHOSUB}
for charged pions. The asymmetry is roughly constant 
at about $0.01$ for $\pi^+$ and compatible with zero for $\pi^-$
in the low and middle range of $z$ and exhibits a steep rise
in the highest $z$ bins for both $\pi^+$ and $\pi^-$. The uncertainties 
from {\sc Pythia} and {\sc rhoMC} Monte Carlo generators are included in
the systematic error band.

\begin{figure}[t]
\vspace*{6.0cm}
\begin{center}
\includegraphics{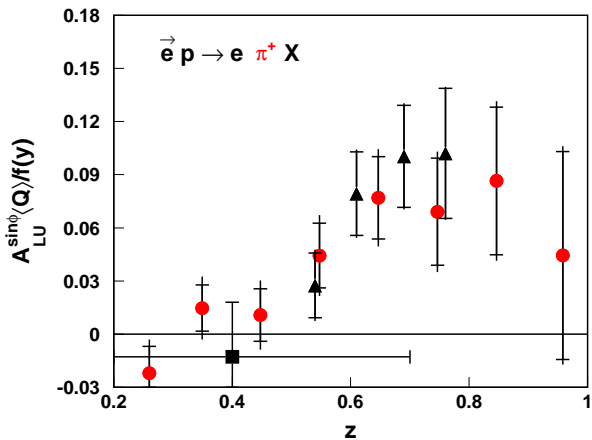}
\caption[*]{Comparison of the kinematically rescaled asymmetry amplitudes
$A^{\sin\phi}_{LU}\cdot\langle Q\rangle/f(\langle y \rangle)$ 
for $\pi^+$  between the \mbox{HERMES} (circles) and
CLAS (triangles)  measurements. 
The full square represents a previous HERMES measurement \cite{AULL}, averaged over the
indicated large $z$ range ($0.2<z<0.7$). The outer error bars represent the
quadratic sum of the systematic uncertainty and the statistical 
uncertainty (inner error bars).
}
\label{HERCLA2}
\end{center}
\end{figure}

A similar measurement \cite{JLAB} has been performed for $\pi^+$ mesons by the
CLAS collaboration at JLab at a  lower beam energy (4.3 GeV),
higher average $x$ (\mbox{$\langle x \rangle \simeq 0.3$}) and lower average
$Q^2$ (\mbox{$\langle Q^2 \rangle \simeq 1.55 $~GeV$^2$)}. 
The comparison  of the \mbox{HERMES} and
CLAS measurements of $A^{\sin\phi}_{LU}$ 
as a function of $z$ is shown for $\pi^+$ in Fig.~\ref{HERCLA2}.
No correction for VM contribution was applied.
In order to account for the different kinematic ranges of the two experiments,
both $A^{\sin\phi}_{LU}$ amplitudes are scaled by 
an $\langle Q \rangle / f(y)$ kinematic prefactor
(cf.~Eq.~\ref{siglu} and \ref{sigmauu}), where
\begin{equation}
f(y)= \frac{y \sqrt{1-y}} {(1-y+y^2/2)}
\label{FYFUN} \quad \quad .
\end{equation}

The $f(y)$ scaled asymmetries
are also often referred to as {\it virtual photon asymmetries} in contrast to the 
usual {\it lepton beam asymmetries}. 
A previous \mbox{HERMES} measurement \cite{AULL} in the range $0.2<z<0.7$  is
added for completeness.
The agreement between the two measurements at different beam energies
indicates that the beam SSA does not exhibit a strong dependence neither on the 
incoming lepton energy nor on $x$.

\begin{figure}[bt]
\vspace*{6.0cm}
\begin{center}
\includegraphics{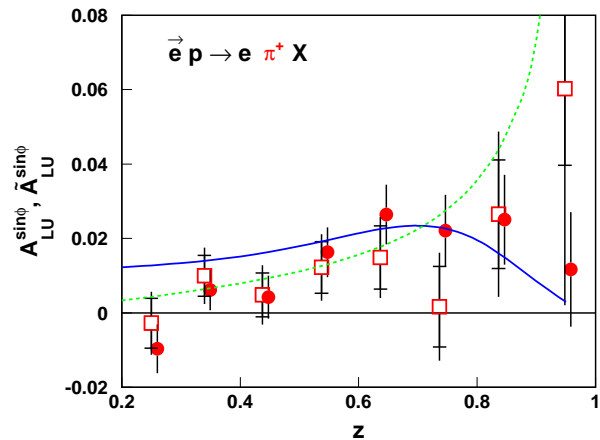}
\caption[*]{Comparison of the $\pi^+$ amplitude uncorrected
  $A_{LU}^{\sin\phi}$ (full circles) and corrected  $\widetilde{A}_{LU}^{\sin\phi}$
  (open squares) for VM contribution 
with model predictions. Outer error bars represent the uncertainties introduced by VM subtraction. A fractional scale uncertainty of $5.5\%$ and a systematic uncertainty of $0.005$ are common. The solid curve represents the  
quark-diquark model \cite{KOGAN}, and the dashed curve the chiral quark model
  \cite{YUAN}. } 
\label{kogan}
\end{center}
\end{figure}

Several theoretical models \cite{KOGAN,YUAN,SCHWEITZER,METZ,AFANAS} have been developed to describe the beam SSA
measurements. 
In Fig.~\ref{kogan} the $A_{LU}^{\sin{\phi}}$ asymmetry amplitudes 
of $\pi^+$ mesons are compared to two model predictions.
The solid curve represents a quark-diquark model calculation \cite{KOGAN},
which claims a dominant contribution from the $eH_1^\perp$ term over the
$h_1^\perp E$ term (cf.~Eq.~\ref{siglu}). In contrast, the 
dashed curve is a calculation using the chiral quark model \cite{YUAN} 
under the assumption that the asymmetry arises solely from the
$h_1^\perp E$ term.  
Both models neglect the contribution of the remaining terms in
Eq.~\ref{siglu}. 
The full circles represent the measurement of the beam SSA for $\pi^+$ mesons
and the empty squares the asymmetry amplitudes $\widetilde{A}_{LU}^{\sin\phi}$
corrected for the contribution from the decay of exclusively produced VMs.
Both models describe 
the data in the low and middle $z$-range equally well,
while in the higher $z$-range, where the influence of the 
exclusive processes in the data increases, the models diverge. 
In the high-$z$ range the quark-diquark
model is in better agreement with the uncorrected asymmetry amplitudes
while the  
corrected asymmetry amplitudes exhibit a rise at 
high $z$ that is described better by the chiral quark model prediction. 
Since the models are not reliable in the high-$z$ range such a comparison
should be treated carefully, and doesn't allow to explicitely rule out 
any of the models with the available statistical accuracy.

In \cite{AFANAS} the $A_{LU}^{\sin(\phi)}$ amplitude is calculated within 
a simple quark-diquark model with a final state gluon exchange without requiring the 
quarks to have spin. There the
contribution of the $g^\perp$ term is considered and the resulting
asymmetry compared to the measurements from \cite{JLAB} and preliminary
HERMES data on asymmetry amplitude dependence on $x$. In addition, 
a pQCD based calculation predicting such asymmetries at a per mille level \cite{AHMED} should be noted.

\begin{table*}[ht]
\caption{\label{aluzdata} Beam SSA: $z$, $x$, and $P_{h\perp}$ dependences of 
 $A_{LU}^{\sin\phi}$ and $\widetilde A_{LU}^{\sin\phi}$ for charged and neutral pions. The systematic uncertainty induced by VM subtraction is given as $\widetilde \Delta_{vm}^{\pi^\pm}$. 
 An additional $5.5\%$ scale uncertainty is due to the beam polarization measurement.
 Other systematic uncertainties do not exceed $0.005$. }
\begin{tabular*}{18.08cm}{c|c|c|c||c|c|c||c|c}
\hline 
$\langle z \rangle$ & $ \langle x\rangle $  & 
$ \langle P_{h\perp}\rangle $ & $\langle Q^2\rangle$ & 
$A_{LU}^{\sin\phi,\pi^+} \pm \Delta_{stat}^{\pi^+}$ & 
$A_{LU}^{\sin\phi,\pi^-} \pm \Delta_{stat}^{\pi^-}$ & 
$A_{LU}^{\sin\phi,\pi^0} \pm \Delta_{stat}^{\pi^0}$ &
$\widetilde A_{LU}^{\sin\phi,\pi^+} \pm \widetilde \Delta_{stat}^{\pi^+} \pm \widetilde \Delta_{vm}^{\pi^+}$&
$\widetilde A_{LU}^{\sin\phi,\pi^-} \pm \widetilde \Delta_{stat}^{\pi^-} \pm \widetilde \Delta_{vm}^{\pi^-}$ 
 \\
\hline 
   0.26 & 0.065 & 0.40 &2.29 & -0.009 $\pm$  0.006  & -0.001 $\pm$  0.007 & 0.002 $\pm$   0.006 & 0.003 $\pm$  0.006 $\pm$ 0.005 & 0.012 $\pm$  0.007 $\pm$ 0.005 \\ 
   0.34 & 0.082 & 0.45 &2.53 &  0.006 $\pm$  0.005  & -0.007 $\pm$  0.006 & 0.028 $\pm$   0.006 & 0.013 $\pm$  0.005 $\pm$ 0.005 & 0.001 $\pm$  0.006 $\pm$ 0.005 \\
   0.44 & 0.092 & 0.46 &2.55 &  0.004 $\pm$  0.005  & -0.007 $\pm$  0.006 & 0.035 $\pm$   0.008 & 0.005 $\pm$  0.005 $\pm$ 0.005 &-0.007 $\pm$  0.007 $\pm$ 0.005 \\
   0.54 & 0.098 & 0.47 &2.51 &  0.016 $\pm$  0.006  & 0.009  $\pm$  0.008 & 0.025 $\pm$   0.009 & 0.011 $\pm$  0.006 $\pm$ 0.005 & 0.000 $\pm$  0.008 $\pm$ 0.006 \\ 
   0.64 & 0.104 & 0.47 &2.47 &  0.026 $\pm$  0.007  & 0.003  $\pm$  0.009 & 0.028 $\pm$   0.012 & 0.012 $\pm$  0.008 $\pm$ 0.006 &-0.018 $\pm$  0.010 $\pm$ 0.008 \\ 
   0.74 & 0.108 & 0.47 &2.37 &  0.021 $\pm$  0.009  & 0.011  $\pm$  0.011 & 0.020 $\pm$   0.015 & 0.000 $\pm$  0.010 $\pm$ 0.008 &-0.023 $\pm$  0.013 $\pm$ 0.011 \\
   0.84 & 0.117 & 0.46 &2.27 &  0.024 $\pm$  0.010  & 0.013  $\pm$  0.014 & 0.019 $\pm$   0.020 & 0.026 $\pm$  0.013 $\pm$ 0.012 & 0.003 $\pm$  0.019 $\pm$ 0.023 \\
   0.95 & 0.128 & 0.45 &2.20 &  0.011 $\pm$  0.013  & 0.007  $\pm$  0.020 & -0.018$\pm$   0.030 & 0.065 $\pm$  0.019 $\pm$ 0.032 & 0.099 $\pm$  0.033 $\pm$ 0.069 \\
\hline
   0.62 & 0.043 & 0.53 & 1.30 & 0.019 $\pm$ 0.010  & 0.016  $\pm$  0.013 & 0.025 $\pm$   0.018 & 0.002$\pm$ 0.012 $\pm$ 0.022 &-0.007 $\pm$  0.015 $\pm$ 0.024\\
   0.63 & 0.075 & 0.44 & 1.84 & 0.024 $\pm$ 0.006  & 0.017  $\pm$  0.008 & 0.016 $\pm$   0.010 & 0.012$\pm$ 0.007 $\pm$ 0.016 &-0.001 $\pm$  0.009 $\pm$ 0.021\\
   0.62 & 0.137 & 0.42 & 3.19 & 0.020 $\pm$ 0.007  & -0.016 $\pm$  0.009 & 0.030 $\pm$   0.012 & 0.012$\pm$ 0.008 $\pm$ 0.009 &-0.030 $\pm$  0.010 $\pm$ 0.013\\
   0.62 & 0.269 & 0.43 & 6.08 & -0.005$\pm$ 0.015  & 0.018  $\pm$  0.022 & 0.037 $\pm$   0.03  &-0.012$\pm$ 0.016 $\pm$ 0.007 & 0.009 $\pm$  0.023 $\pm$ 0.008 \\
\hline                                                                                                                              
   0.62 & 0.108 & 0.17 & 2.44 & 0.019 $\pm$ 0.009  &  0.004 $\pm$  0.01   & 0.000$\pm$   0.015 & 0.001$\pm$      0.009 $\pm$ 0.021 &-0.025 $\pm$  0.012 $\pm$ 0.031\\
   0.62 & 0.105 & 0.35 & 2.45 & 0.028 $\pm$ 0.007  &  0.028 $\pm$  0.009  & 0.035$\pm$   0.012 & 0.013$\pm$ 0.008 $\pm$ 0.020 & 0.006 $\pm$  0.010 $\pm$ 0.030 \\
   0.62 & 0.103 & 0.54 & 2.52 & 0.027 $\pm$ 0.008  &  0.012 $\pm$  0.010  & 0.049$\pm$   0.013 & 0.024$\pm$ 0.009 $\pm$ 0.009 & 0.004 $\pm$  0.012 $\pm$ 0.011\\
   0.61 & 0.095 & 0.74 & 2.49 & 0.015 $\pm$ 0.011  & -0.030 $\pm$  0.016 & 0.009 $\pm$   0.018 & 0.015$\pm$ 0.012 $\pm$ 0.006 &-0.036 $\pm$  0.017 $\pm$ 0.006\\
   0.61 & 0.084 & 1.02 & 2.39 & -0.009$\pm$ 0.014  & -0.054 $\pm$  0.020 & 0.005 $\pm$   0.020 &-0.010$\pm$ 0.014 $\pm$ 0.005 &-0.058 $\pm$  0.020 $\pm$ 0.005\\
\hline

\end{tabular*}
\end{table*}

In conclusion, beam-spin asymmetry amplitudes $A_{LU}^{\sin\phi}$ have been
measured in the azimuthal distribution of pions produced
in semi-inclusive DIS. The 
amplitudes $A_{LU}^{\sin\phi}$
for $\pi^+$ are consistent with zero for low $z$
and exhibit a rise with increasing $z$, reaching values of about $0.02$.
This result is in reasonable agreement with calculations based on
quark-diquark and chiral quark models, as well as with data measured previously
at different kinematics. The comparison with the latter
gives no experimental evidence for a strong dependence of the amplitudes $A_{LU}^{\sin\phi}$ on the beam energy.

The results of the first measurement of $A_{LU}^{\sin\phi}$ for 
negative pions are consistent with zero within the available statistical
accuracy. 
The asymmetry for neutral pions is roughly constant and of the order of
$0.03$, decreasing to zero only
in the lowest and highest $z$ regions.

The variety of components in the polarized cross-section (c.f.~Eq.~\ref{siglu})
make it difficult to disentangle them and find a dominant source of the 
beam-spin asymmetry. Two theoretical models which both describe the data well
are based on opposite assumptions. 
Other models which do not
include the quark spin degrees of freedom predict sizeable asymmetries.
It would be interesting to extend these studies
to double-hadron production integrated over the transverse momentum of
the hadron pair as there all terms in Eq.~\ref{siglu} that explicitely require
intrinsic transverse momentum vanish \cite{Bacchetta:2003vn}.

The contribution
of exclusive VM production to the SIDIS asymmetries observed indicates
that a proper investigation of the hadron production process through
fragmentation requires a
good knowledge of 
exclusive meson production mechanisms and corresponding asymmetries. The
latter  
have been investigated using experimental data and Monte-Carlo
simulations. They were used to obtain the asymmetry amplitudes for 
pions produced in the fragmentation process in SIDIS 
without contamination by vector mesons. 

In a global analysis the beam SSA combined with 
longitudinal and transverse target spin asymmetries in single- and double-hadron
production, as well as jet asymmetries,
if available, will allow a more complete understanding of the nucleon structure as well
as the underlying mechanisms of the quark fragmentation.

\begin{acknowledgments} 
We gratefully acknowledge the DESY management for its support and the staff
at DESY and the collaborating institutions for their significant effort.
This work was supported by the FWO-Flanders, Belgium;
the Natural Sciences and Engineering Research Council of Canada;
the National Natural Science Foundation of China;
the Alexander von Humboldt Stiftung;
the German Bundesministerium f\"ur Bildung und Forschung (BMBF);
the Deutsche Forschungsgemeinschaft (DFG);
the Italian Istituto Nazionale di Fisica Nucleare (INFN);
the MEXT, JSPS, and COE21 of Japan;
the Dutch Foundation for Fundamenteel Onderzoek der Materie (FOM);
the U. K. Engineering and Physical Sciences Research Council, the
Particle Physics and Astronomy Research Council and the
Scottish Universities Physics Alliance;
the U. S. Department of Energy (DOE) and the National Science Foundation (NSF);
the Russian Academy of Science and the Russian Federal Agency for 
Science and Innovations;
the Ministry of Trade and Economical Development and the Ministry
of Education and Science of Armenia;
and the European Community-Research Infrastructure Activity under the
FP6 ''Structuring the European Research Area'' program
(HadronPhysics, contract number RII3-CT-2004-506078).
\end{acknowledgments}

\end{document}